\documentclass{article}

\usepackage{arxiv}

\usepackage[utf8]{inputenc}      
\usepackage[T1]{fontenc}         
\usepackage[hidelinks]{hyperref} 
\usepackage{url}                 
\usepackage{booktabs}            
\usepackage{amsfonts}            
\usepackage{nicefrac}            
\usepackage{microtype}           
\usepackage{lipsum}
\usepackage{graphicx}
\graphicspath{ {./images/} }

\usepackage{orcidlink}
\usepackage{amsmath}
\usepackage{bm}
\usepackage{subcaption}
\usepackage{booktabs}
\usepackage{multirow}
\usepackage{xcolor}
\usepackage{listings}
\usepackage{cite}

\providecommand{\lstbasicfont}{\ttfamily}

\newcommand{\authorcontributions}[1]{%
\vspace{6pt}\noindent{\fontsize{9}{11.2}\selectfont\textbf{Author Contributions:} {#1}\par}}

\title{Implementation and verification of the resolved Reynolds stress transport equations in OpenFOAM}

\author
{
Mario J. Rincón$^{1,2,*}$\orcidlink{0000-0003-3239-6612},
Christoffer Hansen$^1$\orcidlink{0009-0006-5110-3917},
Martino Reclari$^2$,
Mahdi Abkar$^{1,*}$\orcidlink{0000-0002-6220-870X}\\[0.6em]
\textit{$^1$Department of Mechanical and Production Engineering, Aarhus University, 8200 Aarhus N, Denmark}\\[0.6em]
\textit{$^2$Quality \& Sustainability Department, Kamstrup A/S, 8660 Skanderborg, Denmark}
}

\begin{document}
\maketitle


\begin{abstract}
The analysis of the Reynolds Stress Transport Equation (RSTE) provides fundamental physical insights that are essential for the development and validation of advanced turbulence models. However, a comprehensive and validated tool for computing the complete RSTE budget is absent in the widely-used open-source Computational Fluid Dynamics (CFD) framework, OpenFOAM. This work addresses this gap by presenting the implementation and \textit{a posteriori} validation of a function object library for calculating all terms of the resolved RSTE budget in Large-Eddy Simulations (LES).
The library is applied to simulate two canonical wall-bounded turbulent flows: a channel flow and a pipe flow, both at a friction Reynolds number of Re$_{\tau}=180$. The implementation is validated through a mesh refinement study where the results from the LES simulations are systematically compared against high-fidelity Direct Numerical Simulation (DNS) data. 
The computed budget terms are observed to converge systematically towards the DNS reference data.
This validation demonstrates that the implemented library accurately captures the intricate balance of all budget terms. 
This contribution provides the open-source CFD community with a powerful utility for detailed turbulence analysis, thereby facilitating deeper physical understanding and accelerating the development of next-generation turbulence models.
\end{abstract}

\section{Introduction}

The accurate prediction of turbulent flows remains a cornerstone of modern fluid dynamics, with profound implications for a multitude of scientific and engineering applications \cite{lesieur1987turbulence, pope2000turbulent}. While high-fidelity approaches such as Direct Numerical Simulation (DNS) and Large-Eddy Simulation (LES) offer detailed insights into turbulence physics, their prohibitive computational cost restricts their use to academic studies at low Reynolds numbers \cite{moin1998direct, sagaut2005large}. Consequently, Reynolds-Averaged Navier-Stokes (RANS) models continue to be the predominant tool for industrial Computational Fluid Dynamics (CFD) due to their computational efficiency and robustness \cite{slotnick2014cfd, menter2021overview}.

The accuracy of RANS simulations is fundamentally dependent on the fidelity of the turbulence model used to approximate the Reynolds Stress Tensor (RST). The majority of widely-used RANS models are linear eddy-viscosity models, which rely on the Boussinesq hypothesis to relate the RST to the mean rate of strain \cite{boussinesq1877essai}. However, the assumption of isotropy inherent in this hypothesis is known to be invalid in a wide range of complex flows, including those with strong streamline curvature, secondary flows, or significant body forces, leading to predictive inaccuracies \cite{speziale1990analytical, gatski2004constitutive}. To improve turbulence models and gain a deeper physical understanding of complex flow phenomena, an analysis of the transport equation for the Reynolds stresses is highly valuable \cite{launder1975progress}. The exact transport equation for the RST, which is called the Reynolds Stress Transport Equation (RSTE) here, but which is also often referred to as the Reynolds stress budget, describes the evolution of the individual stress components. Each term within this equation---such as production, dissipation, pressure-strain correlation, and turbulent and viscous diffusion---represents a distinct physical mechanism responsible for the creation, destruction, and redistribution of the Reynolds stresses. A term-by-term analysis of this budget therefore provides crucial information that is essential for the development and validation of more sophisticated turbulence closures, including second-moment closure models and advanced data-driven approaches \cite{DURAISAMY2025311, launder1989second, rincon2023progressive, amarloo2023progressive,
hansen2023pod,hansen2024extension,rincon2025generalisable}.

Despite its diagnostic power, the detailed evaluation of the RST budget is not a standard feature in many general-purpose CFD software packages. In particular, for the widely-used open-source CFD framework OpenFOAM \cite{weller1998tensorial}, no readily available and validated utility for a comprehensive RST budget analysis exists. This limitation presents a significant barrier to researchers and engineers who require detailed turbulence analysis for model development or for understanding the intricate physics of their specific applications. The implementation of such a tool is non-trivial, demanding careful numerical treatment to ensure accuracy and consistency with the flow solver. To address this critical gap, a library for the calculation of all terms in the resolved RSTE in LES has been developed and implemented within the OpenFOAM framework. This new functionality enables a complete decomposition of the resolved RST budget, providing a powerful diagnostic tool for the turbulence modelling community. However, the development of the code alone is insufficient; its credibility relies upon rigorous verification and validation. Therefore, a primary contribution of this work is the comprehensive assessment of the implemented library against benchmark data of the highest available fidelity.

The verification and validation of the developed tool have been performed against high-quality DNS data from studies of canonical turbulent flows \cite{el2013direct, lee2015direct}. By comparing the resolved RST budget terms computed from OpenFOAM-based highly resolved LES with the reference DNS data, a term-by-term quantification of the implementation's accuracy is achieved. This process ensures that the library provides a reliable and accurate representation of the resolved RST budget, thereby establishing it as a trustworthy tool for scientific research and advanced engineering analysis.

This study is organised as follows. The governing equations and the theoretical formulation of the resolved RSTE are presented in Section~\ref{sec:RSTE}. The numerical implementation of the budget terms as a function object library in OpenFOAM is detailed in Section~\ref{sec:implementation}. In Section~\ref{sec:results}, the results of the verification and validation studies are presented, where the computed budget terms are compared against DNS data for two benchmark cases. Finally, concluding remarks and an outlook on future work are provided in Section~\ref{sec:conclusions}.

\section{Resolved Reynolds stress transport equations}
\label{sec:RSTE}

This section gives an introduction to the resolved RSTE in LES.
First, the general form is covered in~\ref{subsec:RSTE_general_form}, including the Turbulent Kinetic Energy (TKE) budget.
After this, the resolved RSTE formulation in Cartesian coordinates is presented in~\ref{subsec:cartesian_coordinates}.
Finally, a short discussion of alternative coordinate systems is given in~\ref{subsec:alternative_coordinate_systems}.

\subsection{General form}
\label{subsec:RSTE_general_form}

The total RSTE for the full velocity and pressure fields can be derived from the Navier-Stokes equations~\cite{pope2000turbulent}.
Here, the focus is on the resolved RSTE which can be derived similarly starting from the filtered Navier-Stokes equations~\cite{meneveau1994statistics}.
In general index notation, i.e., without specifying a coordinate system, the resolved RSTE can be written as
\begin{align}
\label{eq:RSTE_general}
    \frac{\partial R_{ij}}{\partial t}
    =
    C_{ij} + P_{ij} + T_{ij} + D_{p,ij} + \Phi_{ij} + D_{\nu,ij} + \varepsilon_{ij} + \Pi_{ij},
\end{align}
where $R_{ij}$ is the resolved RST.
The remaining terms are: $C_{ij}$ convection, $P_{ij}$ production, $T_{ij}$ turbulent transport, $D_{p,ij}$ pressure-diffusion, $\Phi_{ij}$ pressure-strain, $D_{\nu,ij}$ viscous diffusion, $\varepsilon_{ij}$ viscous dissipation, $\Pi_{ij}$ subgrid-scale (SGS) term.
Beyond the individual components of the resolved RSTE budget, the TKE budget is often a primary quantity of interest.
It is obtained as half the trace of Eq.~\eqref{eq:RSTE_general} and reads
\begin{align}
    \frac{\partial k}{\partial t}
    =
    C_k + P_k + T_k + D_{p,k} + D_{\nu,k} + \varepsilon_k + \Pi_k ,
\end{align}
where $k = \tfrac{1}{2} \text{tr}(R_{ij})$ is the resolved TKE, $C_k = \tfrac{1}{2} \text{tr}(C_{ij})$ is the TKE convection, etc.
It should be noted that the pressure-strain term is absent from the TKE budget as it vanishes due to incompressibility; $\Phi_k = \tfrac{1}{2} \text{tr}(\Phi_{ij}) = \langle \tilde{p}^{\prime} \partial_j \tilde{u}^{\prime}_j \rangle = 0$.
Here, $\tilde{\, \cdot \,}$ indicates a low pass filtering operation.

In this study, LES using eddy viscosity based SGS models are considered.
Therefore, the deviatoric part of the SGS tensor $\tau_{ij}$ is modelled as
\begin{align}
    \tau_{ij}^d
    =
    \tau_{ij} - \frac{1}{3} \tau_{kk} \delta_{ij}  
    =
    - 2 \nu_t \tilde{S}_{ij},
\end{align}
where $\tilde{S}_{ij}$ is the filtered rate-of-strain tensor and $\nu_t$ is the eddy viscosity.
In LES, this deviatoric part is included in the filtered momentum equations, while the isotropic part is absorbed into the a modified pressure $p^* = p - \tfrac{1}{3} \tau_{kk}$.
In remainder of this study, it is understood to be that the pressure and SGS tensor used in the resolved RSTE are the modifield pressure and deviatoric part of the SGS tensor.
Similarly, for simplicity, the asterisk superscript on the modified pressure is not written explicitly.

\subsection{Cartesian coordinates}
\label{subsec:cartesian_coordinates}

In Cartesian coordinates and using the Einstein summation convention, the terms on the right-hand side of the resolved RSTE in Eq.~\eqref{eq:RSTE_general} take the following form (see, e.g.,~\cite{meneveau1994statistics})
\begin{align}
    C_{ij} &= -\langle \tilde{u}_k \rangle \frac{\partial \langle \tilde{u}^{\prime}_i \tilde{u}^{\prime}_j \rangle}{\partial x_k},
    \\
    P_{ij} &= - \left( \langle \tilde{u}^{\prime}_j \tilde{u}^{\prime}_k \rangle \frac{\partial \langle \tilde{u}_i \rangle}{\partial x_k} + \langle \tilde{u}^{\prime}_i \tilde{u}^{\prime}_k \rangle \frac{\partial \langle \tilde{u}_j \rangle}{\partial x_k} \right),
    \\
    T_{ij} &= - \frac{\partial \langle \tilde{u}^{\prime}_i \tilde{u}^{\prime}_j \tilde{u}^{\prime}_k \rangle}{\partial x_k} ,
    \\
    D_{p,ij} &= - \left( \frac{\partial \langle \tilde{u}^{\prime}_i \tilde{p}^{\prime} \rangle}{\partial x_j} + \frac{\partial \langle \tilde{u}^{\prime}_j \tilde{p}^{\prime} \rangle}{\partial x_i} \right),
    \\
    \Phi_{ij} &= \left\langle \tilde{p}^{\prime} \left( \frac{\partial \tilde{u}^{\prime}_i}{\partial x_j} + \frac{\partial \tilde{u}^{\prime}_j}{\partial x_i} \right) \right\rangle,
    \\
    D_{\nu,ij} &= \nu \frac{\partial^2 \langle \tilde{u}^{\prime}_i \tilde{u}^{\prime}_j \rangle}{\partial x_k \partial x_k},
    \\
    \varepsilon_{ij} &= -2\nu \left\langle \frac{\partial \tilde{u}^{\prime}_i}{\partial x_k} \frac{\partial \tilde{u}^{\prime}_j}{\partial x_k} \right\rangle,
    \\
    \Pi_{ij} &= -  \left( \left\langle \tilde{u}^{\prime}_j  \frac{ \partial \tau_{ik}^d}{\partial_k} \right\rangle
    + \left\langle \tilde{u}^{\prime}_i \frac{\partial \tau_{jk}^d}{\partial_k} \right\rangle \right).
\end{align}
Here, $\tilde{u}_i$ is the filtered velocity, $\tilde{p}$ is the filtered modified kinematic pressure, $\tau_{ij}^d$ is the deviatoric part of the SGS tensor, $\nu$ is the kinematic viscosity, $\langle \, \cdot \, \rangle$ is an expectation operation taken as time-averaging in this study, and the fluctuating variables are defined as $\tilde{f}' = \tilde{f} - \langle \tilde{f} \rangle$.
Note that this representation of the SGS term used here differs from~\cite{meneveau1994statistics}, where it is decomposed into two parts, however, in this study, the full SGS contribution is kept as a single term.
The explicit SGS term used here in the resolved RSTE can be written as
\begin{align}
    \Pi_{ij}
    =
    2 \left( \left\langle \tilde{u}^{\prime}_j \frac{\partial (\nu_t \tilde{S}_{ik})}{\partial_k} \right\rangle + \left\langle \tilde{u}^{\prime}_i \frac{\partial (\nu_t \tilde{S}_{jk})}{\partial_k}  \right\rangle \right) .
\end{align}

Before moving on, it is important to highlight that because OpenFOAM uses a Cartesian coordinate system, all derivative operations needed for calculating the terms in the Cartesian RSTE budget can be evaluated directly using OpenFOAM functionality.
However, several additional fields, which are not native to OpenFOAM, still need to be added and averaged.
Further discussion of these and other implementation details are given in Section~\ref{sec:implementation}.

\subsection{Alternative coordinate systems}
\label{subsec:alternative_coordinate_systems}

While the Cartesian formulation of the RSTE discussed in~\ref{subsec:cartesian_coordinates} above is the one most commonly used, the RSTE can in principle be derived in any coordinate system.
This can be done by deriving the RSTE in general tensor form, valid for arbitrary curvilinear coordinate systems, using tools from tensor calculus~\cite{moser1984direct}.
The particular form for a given coordinate system then follows directly by specifying the relevant geometric and differential quantities, e.g., the metric tensor and the nabla ($\nabla$) operator.
However, the resulting RSTE formulations for non-Cartesian cases are typically much more involved.
Furthermore, in OpenFOAM, everything is required to be calculated using the Cartesian quantities available during the simulation, which adds an additional level of complexity.
The implementation of the RSTE in non-Cartesian coordinate systems is therefore not pursued in this work.
However, while non-Cartesian coordinate systems are not pursued directly, the current implementation still gives access to the TKE budget, which is coordinate-independent (the trace is the first invariant of a second-rank tensor).
This will be illustrated further in \ref{subsec:pipe_flow} for the case of pipe flow.

\section{Implementation details}
\label{sec:implementation}

This section discusses the resolved RSTE implementation details.
Initially, a quick overview of the standard fields available in OpenFOAM and how their means are calculated is provided for reference.
The resolved mean velocity, mean pressure, and Reynolds stresses are defined as follows
\begin{align}
    \langle \tilde{u}_i \rangle , 
    \qquad 
    \langle \tilde{p} \rangle,
    \qquad
    \langle \tilde{u}^{\prime}_i \tilde{u}^{\prime}_j \rangle .
\end{align}
While it is clear how $\langle \tilde{u}_i \rangle$ and $\langle \tilde{p} \rangle$ can be calculated on the fly, however, it is less immediately apparent for the Reynolds stresses and other higher-order statistics.
In order to achieve this, and following the methodology in OpenFOAM, the resolved Reynolds stresses are rewritten as
\begin{align}
    \langle \tilde{u}^{\prime}_i \tilde{u}^{\prime}_j \rangle 
    =
    \langle \tilde{u}_i \tilde{u}_j \rangle - \langle \tilde{u}_i \rangle \langle \tilde{u}_j \rangle ,
\end{align}
and then $\langle \tilde{u}_i \tilde{u}_j \rangle$ are calculated on the fly, similarly to $\langle \tilde{u}_i \rangle$ and $\langle \tilde{u} \rangle$.
The Reynolds stresses $\langle \tilde{u}^{\prime}_i \tilde{u}^{\prime}_j \rangle$ can then be calculated from $\langle \tilde{u}_i \rangle$ and $\langle \tilde{u}_i \tilde{u}_j \rangle$ when needed for write-out or additional calculations.
Beyond the mean velocity and Reynolds stresses, several additional fields are required for the RSTE budget.
The general strategy taken in the implementation is to use built-in OpenFOAM functionality as much as possible.
However, this comes at the cost of the current implementation being somewhat memory inefficient.
Specifically, defining additional fields is done using the \emph{components}, \emph{grad}, and \emph{multiply} functionalities.
The averaging of these new fields is then performed using \emph{fieldAverage}.

Below, each term in the resolved RSTE budget is reviewed, any potential additional fields required are identified, and the calculation of their averaging is discussed.
Likewise, examples of the OpenFOAM code for calculating the different terms are also given to illustrate the implementation.

\subsection{Convective term}
\label{subsec:convective_term}

The resolved convective term is given by
\begin{align}
    C_{ij} = -\langle \tilde{u}_k \rangle \frac{\partial \langle \tilde{u}^{\prime}_i \tilde{u}^{\prime}_j \rangle}{\partial x_k} .
\end{align}
For the implementation, the convective term is written in conservative form as follows
\begin{align}
    C_{ij} 
    =
    - \frac{\partial}{\partial x_k} \left( \langle \tilde{u}_k \rangle \langle \tilde{u}^{\prime}_i \tilde{u}^{\prime}_j \rangle \right) .
\end{align}
The implementation then boils down to calculating the vectors ($k$ being the vector index)
\begin{align}
    \langle \tilde{u}_k \rangle \langle \tilde{u}^{\prime}_i \tilde{u}^{\prime}_j \rangle ,
\end{align}
for all $i$ and $j$ in the upper diagonal, taking the divergence of these vectors using the \emph{div} functionality, and then collecting them appropriately in a \emph{volSymmTensorField}.
The OpenFOAM code for calculating $C_{11}$ (denoted $C_{xx}$ in the code) is given below as an example:

\begin{lstlisting}[emph={ddt,div,laplacian}]
    // ---------- Calculate Cxx component ---------- //     

    // Define field to store vector
    volVectorField u_uu
    (
        IOobject
        (
            "u_uu_tmp", 
            mesh().time().timeName(), 
            mesh(), 
            IOobject::NO_READ, 
            IOobject::NO_WRITE
        ),
        mesh(), 
        dimensionedVector
        (
            "zero", 
            dimVelocity*dimVelocity*dimVelocity,
            vector::zero
        )
    );
    
    // Calculate vector 
    // Here u, v, and w are vel. comps. and uu is the xx Reynolds stress comp.
    vectorField& resultxx = u_uu.ref();
    forAll(resultxx, i)
    {
        resultxx[i].x() = u[i] * uu[i];
        resultxx[i].y() = v[i] * uu[i];
        resultxx[i].z() = w[i] * uu[i];
    }
    u_uu.correctBoundaryConditions();
    
    // Calculate divergence                                                
    volScalarField Cxx
    (
        IOobject
        (
            "Cxx", 
            mesh().time().timeName(), 
            mesh(), 
            IOobject::NO_READ, 
            IOobject::AUTO_WRITE
        ),
        fvc::div(u_uu)
    );
    Cxx.correctBoundaryConditions();
\end{lstlisting}

\subsection{Production term}
\label{subsubsec:production_term}

The resolved production term is given by
\begin{align}
    P_{ij} = - \left( \langle \tilde{u}^{\prime}_j \tilde{u}^{\prime}_k \rangle \frac{\partial \langle \tilde{u}_i \rangle}{\partial x_k} + \langle \tilde{u}^{\prime}_i \tilde{u}^{\prime}_k  \rangle \frac{\partial \langle \tilde{u}_j \rangle}{\partial x_k} \right) .
\end{align}
For the implementation, the velocity gradient tensor is first calculated from the mean velocity using \emph{grad} and is then contracted with the Reynolds stress tensor.
The resulting tensor is then made symmetric using the \emph{symm} functionality, and the negative sign is added.
Note that \emph{symm} includes a factor of a half which is compesentated by multiplying the result by two.
The OpenFOAM code for calculating the full production term is given below:

\begin{lstlisting}[emph={ddt,div,laplacian}]
    // ---------- Calculate production term ---------- //

    // Calculate velocity gradient tensor
    volTensorField gradUMean = fvc::grad(UMean);

    // Calculate production 
    volSymmTensorField P_ij_RSTE
    (
        IOobject
        (
            prodOutNm, mesh().time().timeName(),
            mesh(), IOobject::NO_READ,
            IOobject::AUTO_WRITE
        ),
        -2.0 * symm(RMean & gradUMean)
    );
    P_ij_RSTE.correctBoundaryConditions();
\end{lstlisting}

\subsection{Turbulent transport term}
\label{subsec:turbulent_transport}

The resolved turbulent transport term is given by
\begin{align}
    T_{ij} = - \frac{\partial \langle \tilde{u}^{\prime}_i \tilde{u}^{\prime}_j \tilde{u}^{\prime}_k \rangle}{\partial x_k} .
\end{align}
Here, the triple products $\langle \tilde{u}^{\prime}_i \tilde{u}^{\prime}_j \tilde{u}^{\prime}_k \rangle$ are additional fields that needs to be calculated and averaged.
The idea is to expand them, similar to the Reynolds stresses, to allow on-the-fly averaging.
Specifically, using $\tilde{u}^{\prime}_i = \tilde{u}_i - \langle \tilde{u}_i \rangle$, the triple correlation can be expanded as
\begin{align}
    \begin{split}
            \langle \tilde{u}^{\prime}_i \tilde{u}^{\prime}_j \tilde{u}^{\prime}_k \rangle 
        &= 
        \langle \tilde{u}_i \tilde{u}_j \tilde{u}_k \rangle 
        - 
        \left[ \langle \tilde{u}_i \rangle \langle \tilde{u}_j \tilde{u}_k \rangle + \langle \tilde{u}_j \rangle \langle \tilde{u}_i \tilde{u}_k \rangle + \langle \tilde{u}_k \rangle \langle \tilde{u}_i \tilde{u}_j \rangle  \right]
        \\ 
        & \ \ \ + 3 \langle \tilde{u}_i \rangle \langle \tilde{u}_j \rangle \langle \tilde{u}_k \rangle  -
        \langle \tilde{u}_i \rangle \langle \tilde{u}_j \rangle \langle \tilde{u}_k \rangle
        \\
        &= 
        \langle \tilde{u}_i \tilde{u}_j \tilde{u}_k \rangle 
        - 
        \left[ \langle \tilde{u}_i \rangle \langle \tilde{u}_j \tilde{u}_k \rangle + \langle \tilde{u}_j \rangle \langle \tilde{u}_i \tilde{u}_k \rangle + \langle \tilde{u}_k \rangle \langle \tilde{u}_i \tilde{u}_j \rangle  \right]
        +
        2 \langle \tilde{u}_i \rangle \langle \tilde{u}_j \rangle \langle \tilde{u}_k \rangle .
    \end{split}
\end{align}
The terms in the square parentheses can be further rewritten using $\tilde{u}_i = \langle \tilde{u}_i \rangle + \tilde{u}^{\prime}_i$ to get
\begin{align}
    \left[ \langle \tilde{u}_i \rangle \langle \tilde{u}_j \tilde{u}_k \rangle + \langle \tilde{u}_j \rangle \langle \tilde{u}_i \tilde{u}_k \rangle + \langle \tilde{u}_k \rangle \langle \tilde{u}_i \tilde{u}_j \rangle  \right]
    =
    3  \langle \tilde{u}_i \rangle \langle \tilde{u}_j \rangle \langle \tilde{u}_k \rangle
    + 
    \left[ \langle \tilde{u}_i \rangle \langle \tilde{u}^{\prime}_j \tilde{u}^{\prime}_k \rangle + \langle \tilde{u}_j \rangle \langle \tilde{u}^{\prime}_i \tilde{u}^{\prime}_k \rangle + \langle \tilde{u}_k \rangle \langle \tilde{u}^{\prime}_i \tilde{u}^{\prime}_j \rangle  \right] .
\end{align}
Putting it all together, the following relation is obtained
\begin{align}
\label{eq:transport_term_implementation}
    \langle \tilde{u}^{\prime}_i \tilde{u}^{\prime}_j \tilde{u}^{\prime}_k \rangle 
    = 
    \langle \tilde{u}_i \tilde{u}_j \tilde{u}_k \rangle  
    - 
    \left[ \langle \tilde{u}_i \rangle \langle \tilde{u}^{\prime}_j \tilde{u}^{\prime}_k \rangle + \langle \tilde{u}_j \rangle \langle \tilde{u}^{\prime}_i \tilde{u}^{\prime}_k \rangle + \langle \tilde{u}_k \rangle \langle \tilde{u}^{\prime}_i \tilde{u}^{\prime}_j \rangle  \right]
    -
    \langle \tilde{u}_i \rangle \langle \tilde{u}_j \rangle \langle \tilde{u}_k \rangle.
\end{align}
Thus, to summarise, the resolved mean velocities $\langle \tilde{u}_i \rangle$, the Reynolds stresses $\langle \tilde{u}^{\prime}_i \tilde{u}^{\prime}_j \rangle$, and the triple-velocity-products $\langle \tilde{u}_i \tilde{u}_j \tilde{u}_k \rangle$ are needed.
The triple-products are constructed by using \emph{components} on the velocity and then using the \emph{multiply} functionality to get the required products.
The averaging is done using \emph{fieldAverage}.
After calculating the fluctuating triple-products $\langle \tilde{u}^{\prime}_i \tilde{u}^{\prime}_j \tilde{u}^{\prime}_k \rangle$ using Eq.~\eqref{eq:transport_term_implementation}, they are collected together in vectors ($k$ is the vector index) for every $i$ and $j$, similar to the convective term.
The divergences of these vectors are then calculated using \emph{div}, the negative sign is added, and, finally, everything is collected in a \emph{volSymmTensorField}.
The OpenFOAM code for calculating $T_{11}$ (denoted $T_{xx}$ in the code) is given below as an example:

\begin{lstlisting}[emph={ddt,div,laplacian}]
    // ---------- Calculate Txx component ---------- //

    // Define field to store vector
    volVectorField UxUxUk
    (
        IOobject
            (
                "UxUxUk_tmp", 
                mesh().time().timeName(), 
                mesh(), IOobject::NO_READ, 
                IOobject::NO_WRITE
            ),
            mesh(), 
            dimensionedVector
            (
                "zero",
                dimVelocity*dimVelocity*dimVelocity,
                vector::zero
            )
    );
                                                
    // Calculate vector (<u'u'u'>, <u'u'v'>, <u'u'w'>)
    vectorField& resultxx = UxUxUk.ref();
    forAll(resultxx, i)
    {
        resultxx[i].x() = ( uuu[i] 
                            - (u[i] * uu[i] + u[i] * uu[i] + u[i] * uu[i]) 
                            - u[i] * u[i] * u[i] );
        resultxx[i].y() = ( uuv[i] 
                            - (u[i] * uv[i] + u[i] * uv[i] + v[i] * uu[i]) 
                            - u[i] * u[i] * v[i] );
        resultxx[i].z() = ( uuw[i] 
                            - (u[i] * uw[i] + u[i] * uw[i] + w[i] * uu[i]) 
                            - u[i] * u[i] * w[i] );
    }
    UxUxUk.correctBoundaryConditions();

    // Calculate divergence
    volScalarField Txx
    (
        IOobject
        (
            outputTxx, 
            mesh().time().timeName(), 
            mesh(), 
            IOobject::NO_READ, 
            IOobject::AUTO_WRITE
        ), 
        fvc::div(UxUxUk)
    );
    Txx.correctBoundaryConditions();
\end{lstlisting}

\subsection{Viscous diffusion term}
\label{subsec:viscous_diffusion}

The resolved viscous diffusion term is given by
\begin{align}
    D_{\nu,ij} = \nu \frac{\partial^2 \langle \tilde{u}^{\prime}_i \tilde{u}^{\prime}_j \rangle}{\partial x_k \partial x_k},
\end{align}
and does not require any new fields.
It only requires the calculation of the Laplacian of the Reynolds stresses using \emph{laplacian}.
However, an explicit calculation of the Laplacian was observed to give spurious results at the first off-wall cell centre.
Therefore, in the implementation, these values are overwritten with the value from the nearest wall-normal neighbour.
Hence, the OpenFOAM code for calculating the full viscous diffusion term is given below: 

\begin{lstlisting}[emph={ddt,div,laplacian}]
    // ---------- Calculate viscous diffusion term ---------- //

    volSymmTensorField D_ij_RSTE
    (
        IOobject
        (
            viscDiffOutNm, mesh().time().timeName(),
            mesh(),
            IOobject::NO_READ,
            IOobject::AUTO_WRITE
        ),
        nu * fvc::laplacian(RMean)
    );
    
    // ---------- Additional code for first cell interpolation ---------- //
    // ... code not included for brevity...
\end{lstlisting}

\subsection{Pressure diffusion term:}
\label{subsec:pressure_diffusion}

The resolved pressure diffusion term is given by
\begin{align}
    D_{p,ij} = 
    - \left( \frac{\partial \langle \tilde{u}^{\prime}_j \tilde{p}^{\prime} \rangle}{\partial x_i} + \frac{\partial \langle \tilde{u}^{\prime}_i \tilde{p}^{\prime} \rangle}{\partial x_j} \right) , 
\end{align}
where it should be remembered that $\tilde{p}$ is the modefied kinematic pressure.
The double-correlations are then expanded as follows
\begin{align}
    \langle \tilde{u}^{\prime}_i \tilde{p}^{\prime} \rangle 
    = 
    \langle \tilde{u}_i \tilde{p} \rangle - \langle \tilde{u}_i \rangle \langle \tilde{p} \rangle .
\end{align}
Next, \emph{multiply} is used to create a new field for the velocity-pressure product and the averaging is done using \emph{fieldAverage}.
The gradient tensor corresponding to $\langle u^{\prime}_i p^{\prime} \rangle$ is calculated, symmetrized using \emph{symm}, and the negative sign is added.
The factor of a half from \emph{symm} is compensated for by multiplying with two.
The OpenFOAM code for calculating the full pressure diffusion term is given below:

\begin{lstlisting}[emph={ddt,div,laplacian}]
    // ---------- Calculate pressure diffusion term ---------- //

    // Calculate fluctuating velocity-pressure product
    volVectorField UPrimePPrimeMean
    (
        IOobject
        (
            "UPrimePPrimeMean",
            mesh().time().timeName(),
            mesh(),
            IOobject::NO_READ,
            IOobject::NO_WRITE
        ),
        UpMean - (UMean * pMean)
    );

    // Calculate corresponding gradient tensor
    volTensorField gradUPrimePPrimeMean = fvc::grad(UPrimePPrimeMean); 

    // Calculate pressure diffusion
    volSymmTensorField Dp_ij_RSTE
    (
        IOobject
        (
            outputDpij,
            mesh().time().timeName(),
            mesh(),
            IOobject::NO_READ,
            IOobject::AUTO_WRITE
        ),
        -2.0 * symm(gradUPrimePPrimeMean)
    );
    Dp_ij_RSTE.correctBoundaryConditions();
\end{lstlisting}

\subsection{Pressure-strain term}
\label{subsec:pressure_strain}

The resolved pressure-strain term is given by
\begin{align}
    \Phi_{ij} 
    = 
    \left\langle \tilde{p}^{\prime} \left( \frac{\partial \tilde{u}^{\prime}_i}{\partial x_j} + \frac{\partial \tilde{u}^{\prime}_j}{\partial x_i} \right) \right\rangle
    =
    2 \langle \tilde{p}^{\prime} \tilde{S}^{\prime}_{ij} \rangle,
\end{align}
where it should be remembered that $\tilde{p}$ is the modefied kinematic pressure and $\tilde{S}^{\prime}_{ij}$ is the filtered fluctuating rate-of-strain tensor.
The double correlation is expanded as follows
\begin{align}
    \langle \tilde{p}^{\prime} \tilde{S}^{\prime}_{ij} \rangle 
    = 
    \langle \tilde{p} \tilde{S}_{ij} \rangle - \langle \tilde{p} \rangle \langle \tilde{S}_{ij} \rangle .
\end{align}
In terms of implementation, a new field consisting of the product of the pressure and the velocity gradient tensor is introduced using \emph{multiply}.
This field is subsequently averaged using \emph{fieldAverage}.
The remaining calculations are given in the OpenFOAM code below:

\begin{lstlisting}[emph={ddt,div,laplacian}]
    // ---------- Calculate pressure-strain term ---------- //

    // Calculate SijMean from UMean
    volSymmTensorField SijMean = symm(fvc::grad(UMean));

    // Calculate pSijMean from pGradUMean
    volSymmTensorField pSijMean = symm(pGradUMean);

    // Calculate pressure-strain
    volSymmTensorField Phi_ij_RSTE
    (
        IOobject
        (
            Phi_ij_RSTE,
            mesh().time().timeName(),
            mesh(),
            IOobject::NO_READ,
            IOobject::AUTO_WRITE
        ),
        2.0 * (pSijMean - pMean * SijMean)
    );
    Phi_ij_RSTE.correctBoundaryConditions();
\end{lstlisting}

\subsection{Dissipation term}
\label{subsec:dissipation}

The resolved dissipation term is given by
\begin{align}
    \varepsilon_{ij} = 2\nu \left\langle \frac{\partial \tilde{u}^{\prime}_i}{\partial x_k} \frac{\partial \tilde{u}^{\prime}_j}{\partial x_k} \right\rangle .
\end{align}
The double correlations are then expanded as
\begin{align}
    \begin{split}
            \left\langle \frac{\partial \tilde{u}^{\prime}_i}{\partial x_k} \frac{\partial \tilde{u}^{\prime}_j}{\partial x_k} \right\rangle
            =
            \left\langle \frac{\partial \tilde{u}_i}{\partial x_k} \frac{\partial \tilde{u}_j}{\partial x_k} \right\rangle
            -
            \frac{\partial \langle \tilde{u}_i \rangle}{\partial x_k} \frac{\partial \langle \tilde{u}_j \rangle}{\partial x_k} .
    \end{split}
\end{align}
Thus, there is a need to introduce new fields for the velocity gradient products. 
This is done using \emph{grad} to get the velocity gradients and \emph{multiply} to construct the products. The averages are then calculated using \emph{fieldAverage}.
The implementation then primarily involves summing up all the scalar contributions.
The OpenFOAM code is given below:

\begin{lstlisting}[emph={ddt,div,laplacian}]
    // ---------- Calculate dissipation term ---------- //

    // Define volSymmTensorField
    volSymmTensorField eps_ij
    (
        IOobject
        (
            outputeps_ij,
            mesh().time().timeName(),
            mesh(),
            IOobject::NO_READ,
            IOobject::AUTO_WRITE
        ),
        mesh(), 
        dimensionedSymmTensor
        (
            "zero",
            dimensionSet(0 ,2 ,-3 ,0 ,0 ,0 ,0), 
            symmTensor::zero
        )
    );

    // Calculate dissipation
    symmTensorField& eps_ij_Field = eps_ij.ref();
    forAll(eps_ij_Field, i)
    {
        
        // Only eps_xx calculation for illustration
        eps_ij_Field[i].xx() = ( 2.0 * nu.value() * 
                                   ( (dUxxdUxx[i] + dUyxdUyx[i] + dUzxdUzx[i]) 
                                     - ( gradUxx[i] * gradUxx[i] 
                                       + gradUyx[i] * gradUyx[i] 
                                       + gradUzx[i] * gradUzx[i] ) 
                                    ) 
                                );

        // Remaining components
        // ... not included for brevity...
    }
    eps_ij.correctBoundaryConditions();
\end{lstlisting}

\subsection{SGS term}
\label{subsec:SGS_term}

The SGS term for an eddy viscosity model is given by
\begin{align}
    \Pi_{ij}
    =
    2 \left( \left\langle \tilde{u}^{\prime}_j \frac{\partial (\nu_t \tilde{S}_{ik}) }{\partial_k} \right\rangle + \left\langle \tilde{u}^{\prime}_i \frac{\partial (\nu_t \tilde{S}_{jk})}{\partial_k}  \right\rangle \right) .
\end{align}
Then, using $\tilde{u}^{\prime}_i = \tilde{u}_i - \langle \tilde{u}_i \rangle$, the double-correlations are decomposed as follows
\begin{align}
    \Pi_{ij}^*
    =
    2\left( \left\langle \tilde{u}_j \frac{\partial (\nu_t \tilde{S}_{ik})}{\partial_k}  \right\rangle + \left\langle \tilde{u}_i \frac{\partial (\nu_t \tilde{S}_{jk})}{\partial_k}  \right\rangle  \right)
    -
    2\left( \langle \tilde{u}_j \rangle \left\langle \frac{\partial (\nu_t \tilde{S}_{ik})}{\partial_k} \right\rangle + \langle \tilde{u}_i \rangle \left\langle \frac{\partial (\nu_t \tilde{S}_{jk})}{\partial_k} \right\rangle \right).
\end{align}
Thus, several new fields need to be introduced.
Further, the divergences $\partial (\nu_t \tilde{S}_{ik}) / \partial_k$ cannot be calculated directly using built-in OpenFOAM functionality.
Therefore, a custum coded function object is made for this purpose. 
In full, $\partial (\nu_t \tilde{S}_{ik}) / \partial_k$ is constructed by first calculating the velocity gradient tensor using \emph{grad} and then multiplied with $\nu_t$ using \emph{multiply}.
The coded function object then symmetrizes this product using \emph{symm} and calculates the divergence using \emph{div}.
Next, the products $\tilde{u}_j \partial (\nu_t \tilde{S}_{ik}) / \partial_k$ are constructed using \emph{multiply} on the velocity components and $\partial (\nu_t \tilde{S}_{ij}) / \partial_k$.
These fields, i.e., $\partial (\nu_t \tilde{S}_{ik}) / \partial_k$ and $\tilde{u}_j \partial (\nu_t \tilde{S}_{ik}) / \partial_k$, are then averaged using \emph{fieldAverage}.
The products $\langle \tilde{u}_j \rangle \langle \partial (\nu_t \tilde{S}_{ik} ) / \partial_k \rangle$ are also constructed using \emph{multiply}.
This gives all the components needed for calculating the SGS term.
The OpenFOAM code putting all the components together is given below:

\begin{lstlisting}[emph={ddt,div,laplacian}]
    // ---------- Calculate SGS term ---------- //

    // Constructing SGS as a volSymmTensorField 
    volSymmTensorField SGS_ij_RSTE
    (
        IOobject
        (
            sgsOutNm, 
            mesh().time().timeName(), 
            mesh(), 
            IOobject::NO_READ, 
            IOobject::AUTO_WRITE
        ),
        2.0 * symm(2.0 * divNutSijUMean) - 2.0 * symm(2.0 * divNutSijMean_UMean)
    );
    SGS_ij_RSTE.correctBoundaryConditions();
\end{lstlisting}

\section{Results}
\label{sec:results}

The resolved RSTE budget implementation is tested on two different wall-bounded turbulent flows: channel and pipe.
An overview of the simulation setups is given in~\ref{subsec:simulation_details}.
This is followed by the channel flow results in~\ref{subsec:channel_flow} and the pipe flow results in~\ref{subsec:pipe_flow}. Since second-order-accurate numerical simulations following the finite volume method are performed in this study, it is not expected that the results fully match the DNS reference data. The reader is reminded that the aim of this study is to show, provide, and validate a seamless and correct calculation of the resolved RSTE budget with OpenFOAM, not to match spectral-high-order data from DNS results.
For brevity and simplicity of the presentation of the results, the "resolved" before, e.g., RSTE, is dropped below together with the $\tilde{\, \cdot \,}$ notation for filtered variables.

\subsection{Simulation details}
\label{subsec:simulation_details}

An overview of the solvers is provided, as well as numerical schemes, SGS modeling, and meshing used in the simulations.
Note that the numerical schemes, SGS modeling, and meshes take inspiration from previous efforts in OpenFOAM-based LES \cite{matai2019large, mukha2019library,nozarian2025laminarization,hansen2026wall}.

The \emph{pimpleFoam} solver based on the PIMPLE algorithm is used with a single outer correction (PISO-like mode), two inner correctors for the pressure-velocity coupling, one non-orthogonal corrector per iteration, and the momentum predictor is enabled.
Time derivatives are discretised using the second-order \emph{backward} scheme, and adaptive time stepping is used to ensure a Courant-Friedrichs-Lewy (CFL) number below $0.5$. 
Spatial gradients and divergence terms are computed with the \emph{Gauss linear} method. 
Viscous terms also use the \emph{Gauss linear} scheme with non-orthogonal correction.
Linear interpolation is used for cell-face values, and surface-normal gradients are corrected for mesh non-orthogonality.
To drive the flow, a momentum source is added using \emph{meanVelocityForce} to enforce a bulk velocity of $U_b = 1$.
The forcing is time-dependent but uniform in space.
To ensure $Re_\tau = 180$ in the simulations, the corresponding bulk Reynolds numbers must be matched.
These are $U_b = 2857$ for channel (from~\cite{lee2015direct}) and $U_b = 5300$ for pipe (from~\cite{el2013direct}), which gives the resulting kinematic viscosities as $\nu = 3.5 \times 10^{-4}$ and $\nu = 1.9 \times 10^{-4}$, respectively.
The momentum equations are solved using a preconditioned biconjugate gradient stabilised (PBiCGStab) solver with a diagonal incomplete LU (DILU) precondition. Furthemore, the linear system for the pressure is solved using a preconditioned conjugate gradient (PCG) solver with a geometric–algebraic multigrid (GAMG) preconditioner and a diagonal incomplete–Cholesky Gauss–Seidel (DICGaussSeidel) smoother with two post-sweeps. Finally, tight absolute numerical tolerances are applied to the final iterations for both velocity and pressure.

For SGS modeling, the Wall-Adapting Local Eddy-viscosity (WALE) model \cite{ducros1998wall}  is used. This SGS model expresses the turbulent eddy-viscosity using the square of the velocity gradient tensor. Denoting the filtered velocity gradient tensor as $\tilde{g}_{ij} =\partial \tilde{u}_j/\partial x_i$, the traceless symmetric part of the square of the velocity gradient tensor is
\begin{align}
    \tilde{\xi}_{ij} = \frac{1}{2} \left( \tilde{g}_{ij}^2 + \tilde{g}_{ji}^2 \right) - \frac{1}{3} \tilde{g}_{kk}^2 ,
\end{align}
where $\tilde{g}_{ij}^2 = \tilde{g}_{ik}\tilde{g}_{kj}$.
The turbulent eddy-viscosity for the WALE model can then be written as
\begin{align}
    \nu_t^{\text{WALE}} = ( C_w \Delta )^2 \frac{( \tilde{\xi}_{ij} \tilde{\xi}_{ij} )^{3/2}}{( \tilde{S}_{ij} \tilde{S}_{ij} )^{5/2} + \left( \tilde{\xi}_{ij} \tilde{\xi}_{ij} \right)^{5/4}} ,
\end{align}
where $C_w$ is a constant, i.e., it is not tuned dynamically.

\begin{figure*}[ht!]
    \centering
    \begin{subfigure}[t]{0.49\textwidth}
        \centering
        \includegraphics[width=\textwidth]{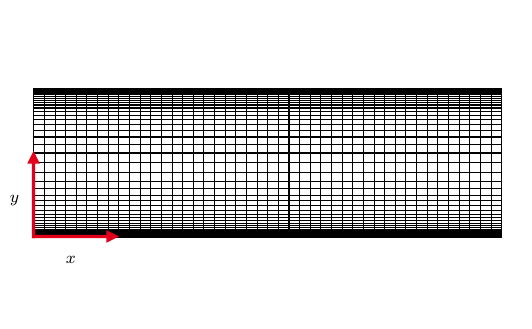}
        \caption{Channel flow mesh.}
        \label{fig:chan_flow_mesh}        
    \end{subfigure}%
    \hfill
    \begin{subfigure}[t]{0.49\textwidth}
        \centering
        \includegraphics[width=\textwidth]{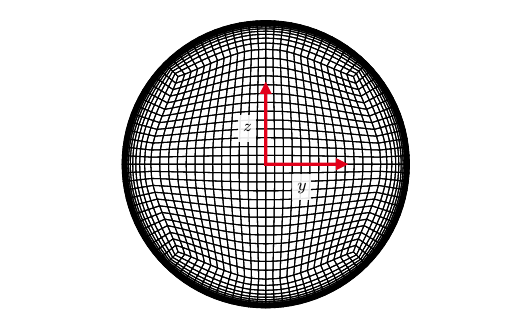}
        \caption{Pipe flow mesh.}
        \label{fig:pipe_flow_mesh}        
    \end{subfigure}
    \caption{Mesh topology for channel (Fig. \ref{fig:chan_flow_mesh}) and pipe flow (Fig. \ref{fig:pipe_flow_mesh}). The number of cells has been reduced in this figure for visualisation purposes.}
    \label{fig:mesh_illustration}
\end{figure*}

For both channel and pipe, structured hexahedral meshes are used.
In both cases, the meshes are close to isotropic in the core of the flow and are then stretched towards the walls to provide sufficient near-wall resolution.
An illustration of the meshes is shown in Fig.~\ref{fig:mesh_illustration}.

To assess the resolution requirements where the RSTE from LES start to faithfully reproduce the full RSTE from DNS, a mesh refinement campaign for both channel and pipe has been performed.
A summary of the meshes used for the different cases is given in Table~\ref{tab:mesh_overview}.

\begingroup
\setlength{\tabcolsep}{8pt} 
\renewcommand{\arraystretch}{1.2} 
\begin{table}[ht!]
\centering
\begin{tabular}{cccccccc}
\toprule
\multirow{2}{*}{\textbf{Channel cases}} & \multirow{2}{*}{$N_\text{tot}$} & \multicolumn{3}{c}{Centre} & \multicolumn{3}{c}{Wall} \\
\cmidrule(lr){3-5} \cmidrule(lr){6-8}
& & $\Delta x_c^+$ & $\Delta y_c^+$ & $\Delta z_c^+$ & $\Delta x_w^+$ & $\Delta y_w^+$ & $\Delta z_w^+$ \\
\midrule
$L_1$ & 131 072   & 17.7 & 17.4 & 17.7 & 17.7 & 0.87 & 17.7 \\
$L_2$ & 389 344   & 12.3 & 13.7 & 12.3 & 12.3 & 0.43 & 12.3 \\
$L_3$ & 1 048 576 & 8.8  & 10.5 & 8.8  & 8.8  & 0.26 & 8.8  \\
$L_4$ & 2 916 000 & 6.3  & 7.9  & 6.3  & 6.3  & 0.16 & 6.3  \\
\midrule
\multirow{2}{*}{\textbf{Pipe cases}} & \multirow{2}{*}{$N_\text{tot}$} & \multicolumn{3}{c}{Centre} & \multicolumn{3}{c}{Wall} \\
\cmidrule(lr){3-5} \cmidrule(lr){6-8}
& & $\Delta z_c^+$ & $\Delta r_c^+$ & $R \Delta \theta_c^+$ & $\Delta z_w^+$ & $\Delta r_w^+$ & $R \Delta \theta_w^+$ \\
\midrule
$L_1$ & 536 640   & 15   & 3.9  & 3.1  & 15   & 0.07 & 2.7  \\
$L_2$ & 1 173 120 & 7.5  & 3.9  & 3.1  & 7.5  & 0.04 & 2.7  \\
$L_3$ & 3 400 800 & 2.7  & 3.9  & 3.1  & 2.7  & 0.02 & 2.7  \\
\bottomrule
\end{tabular}
\vspace{1mm}
\caption{Mesh resolution summary for channel and pipe flow simulations. The table shows the total number of grid points ($N_\text{tot}$) and the dimensionless cell sizes in wall units ($\Delta^+$) for both centre and wall regions. Centre cell sizes represent the mesh resolution in the bulk flow region, while wall cell sizes correspond to the near-wall mesh refinement. The wall-normal direction ($y$ for channel, $r$ for pipe) shows the finest resolution near the walls.}
\label{tab:mesh_overview}
\end{table}
\endgroup

\subsection{Channel Flow}
\label{subsec:channel_flow}

To ensure a correct implementation of the RSTE budget in LES, results for different meshes are compared against DNS data from \cite{lee2015direct} of channel flow at Re$_{\tau}=180$.
This section displays 1D-mean values obtained by performing a stream-span-wise spatial average of the first and second order statistics of the velocity---$\langle u \rangle$, $k$, $R_{ij}$---as well as the RSTE budget terms for the 4 main components of the tensor---$11$, $22$, $33$, $12$.
Here, $1$, $2$, and $3$ refer to the streamwise, wall-normal, and spanwise directions, respectively.
All values are shown in wall units.

\begin{figure}[ht!]
    \centering
    \includegraphics[width=0.95\linewidth]{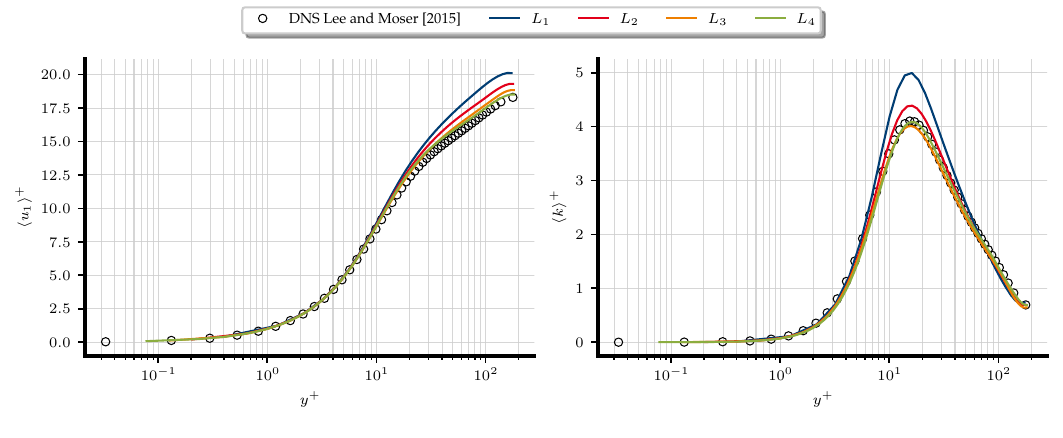}
    \caption{Velocity $\langle u \rangle$ and TKE $k$ profiles.}
    \label{fig:chan_UMean_TKE}
\end{figure}

As seen in Fig. \ref{fig:chan_UMean_TKE}, the turbulent flow profile of the velocity and TKE are predicted with increased accuracy by the finer meshes. Both in magnitude and gradients, meshes with fineness above $L_3$ predict the velocity and TKE with marginal deviations from DNS data and display physical consistency. The mean velocity profile exhibits the characteristic logarithmic region in the range $y^{+} \in [30, 100]$, where the velocity follows the law of the wall. The TKE profile shows a known maximum in the buffer region around $y^{+} \approx 15$; this maximum arises from the intense shear production of turbulence, which is highest in this region, coupled with the onset of viscous dissipation, which becomes dominant closer to the wall. In terms of TKE prediction, the finer meshes capture the near-wall behaviour more accurately, particularly in the buffer layer ($5 < y^{+} < 30$), where gradients are steepest.

\begin{figure}[ht!]
    \centering
    \includegraphics[width=0.95\linewidth]{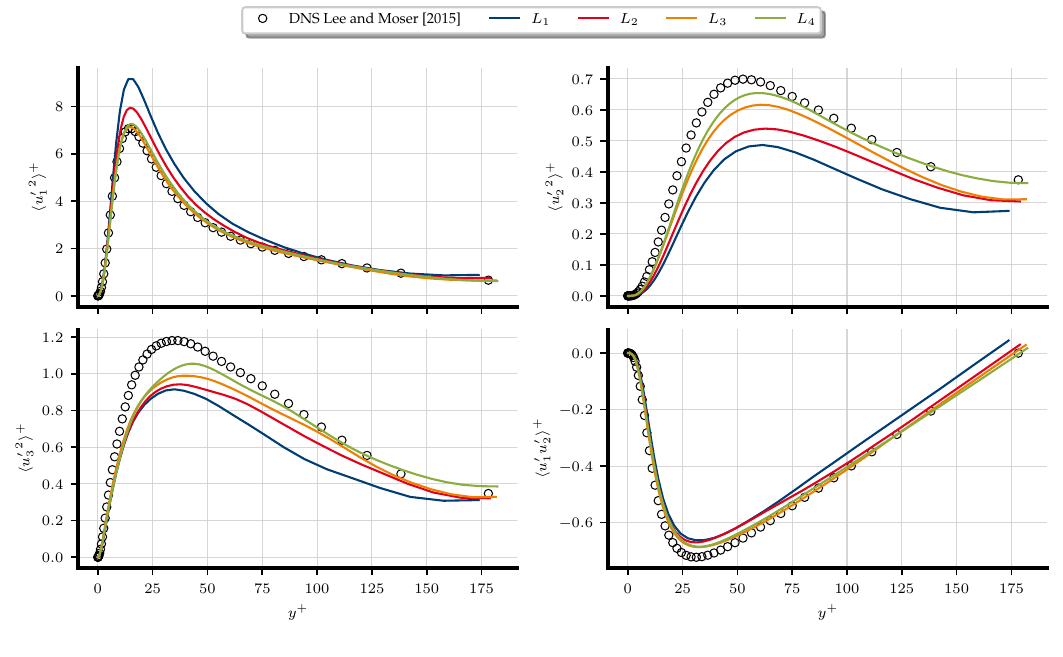}
    \caption{Profiles of $R_{ij}$, specifically components $11$, $22$, $33$, and $12$.}
    \label{fig:chan_UPrime2Mean}
\end{figure}

Similarly, in Fig. \ref{fig:chan_UPrime2Mean}, the $11$, $22$, $33$, and $12$ components of the $R_{ij}$ are predicted with increased accuracy by the finer meshes. Gradients are predicted following DNS data, although the magnitudes of the $22$ and $33$ components do not fully match DNS results at regions $y^{+}\in \left[ 25-75 \right]$. It should be noted that some discrepancies are expected in turbulence simulations when comparing results from solvers with different numerics \cite{chen2023quantifying}.
Nonetheless, these uncertainties associated with numerics do not fully account for observed differences.
Instead, the major contributor to the observed difference is likely the dissipative numerics in OpenFOAM, which is discussed in further detail in \cite{montecchia2019improving}.
The $R_{11}$ component (streamwise normal stress) shows the largest magnitude, peaking in the buffer region due to the strong mean shear production. The $R_{22}$ and $R_{33}$ components (wall-normal and spanwise normal stresses) are smaller in magnitude, reflecting the anisotropic nature of wall-bounded turbulence. The $R_{12}$ component (Reynolds shear stress) is crucial for momentum transport and shows a maximum around $y^{+} \approx 30$, where the production term balances the pressure-strain correlation.

\begin{figure}[ht!]
    \centering
    \includegraphics[width=0.95\linewidth]{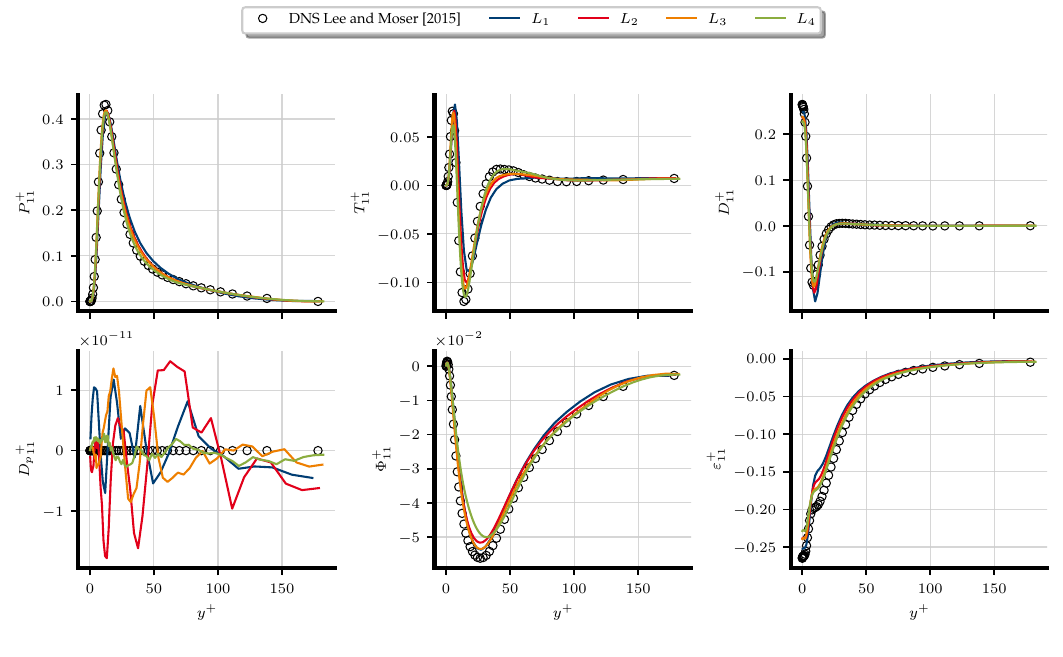}
    \caption{RSTE budget for the streamwise normal stress component $\langle u^{\prime}_1 u^{\prime}_1\rangle$.}
    \label{fig:chan_uu_RSTE}
\end{figure}

Figure \ref{fig:chan_uu_RSTE} presents the RSTE budget for the streamwise component $R_{11}$. The budget terms include production $P^+_{11}$, turbulent transport $T^+_{11}$, viscous diffusion $D^+_{11}$, pressure-strain correlation $\Phi_{11}$, pressure-diffusion $D^{+}_{p, 11}$, and viscous dissipation $\varepsilon^+_{11}$. The production term is the dominant source, converting mean flow energy into turbulent fluctuations. This is balanced primarily by the pressure-strain correlation, which acts as a sink term here by redistributing energy from the streamwise direction to the wall-normal and spanwise components, and by the dissipation term. The turbulent transport and pressure diffusion terms are responsible for spatially redistributing the energy away from the maximum production region near the wall. The simulation predictions for all terms follow the previously seen trends --- results on finer meshes more accurately follow DNS data, and all results show accurate gradients and magnitudes. In the DNS reference data, the pressure-diffusion term is given as zero; however, the simulation data shown is displayed with the remaining numerical fluctuations on the order of $10^{-11}$, which is close to the numerical precision used.

\begin{figure}[ht!]
    \centering
    \includegraphics[width=0.95\linewidth]{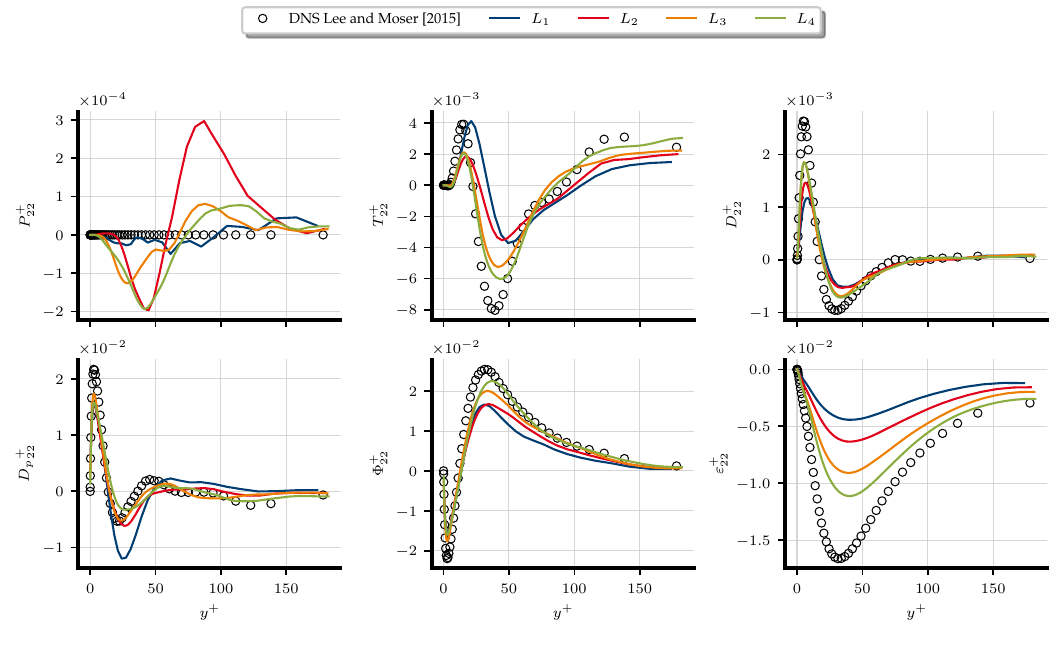}
    \caption{RSTE budget for the wall-normal stress component $\langle u^{\prime}_2 u^{\prime}_2 \rangle$.}
    \label{fig:chan_vv_RSTE}
\end{figure}

The budget for the wall-normal component $R_{22}$ is shown in Fig. \ref{fig:chan_vv_RSTE}. It is important to highlight that $R_{22}$ has no direct production term since the mean wall-normal velocity is zero. Its energy is supplied entirely by the pressure-strain correlation $\Phi_{22}$, which acts as a source term by redirecting the excess energy from the streamwise component. A significant portion of this term is explained by the \textit{wall-reflection} effect, which damps wall-normal fluctuations and enhances in-plane fluctuations. This energy gain is then balanced by dissipation and spatial transport. Once again, finer meshes more accurately predict DNS data and DNS trends are followed in all predictions. The greatest discrepancy in the data is shown at $\varepsilon^{+}_{22}$; nonetheless the data shows that further mesh refinement tends towards the DNS reference. For $P^{+}_{22}$, the numerical fluctuactions are on the order of $10^{-4}$, which are considered acceptable.

\begin{figure}[ht!]
    \centering
    \includegraphics[width=0.95\linewidth]{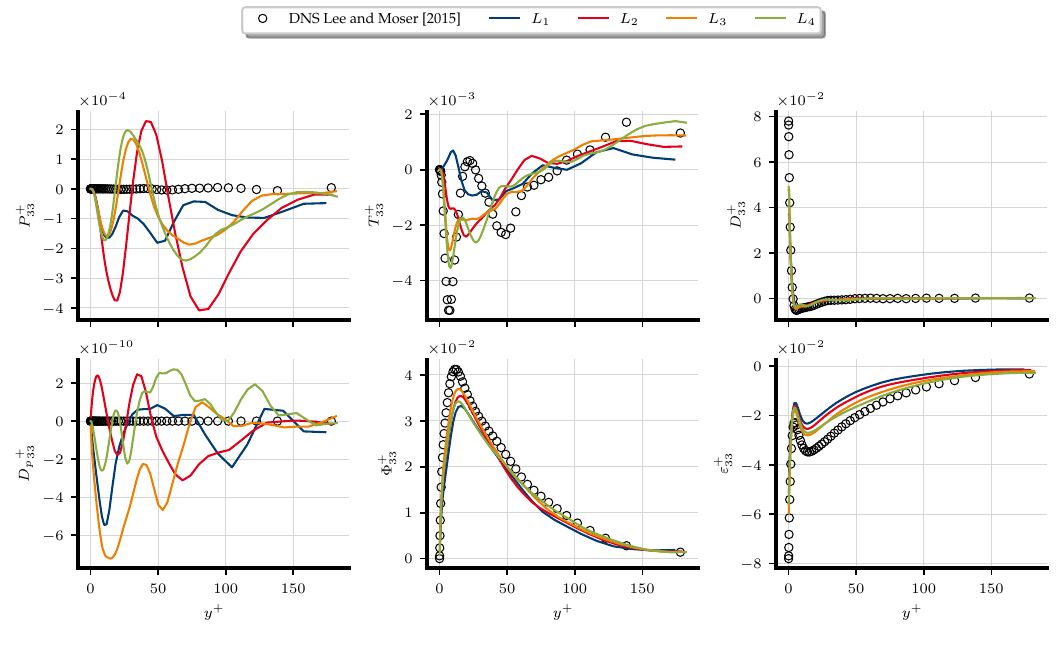}
    \caption{RSTE budget for the spanwise normal stress component $\langle u^{\prime}_3 u^{\prime}_3 \rangle$.}
    \label{fig:chan_ww_RSTE}
\end{figure}

The spanwise component budget in Fig. \ref{fig:chan_ww_RSTE} displays the transport equation for $R_{33}$. Similar to the wall-normal component, it has no direct production and relies on energy redistribution via the pressure-strain correlation $\Phi_{33}$. The spanwise component typically shows the smallest magnitude among the normal stresses, reflecting the quasi-two-dimensional nature of the large-scale structures in wall-bounded turbulence. The pressure-strain correlation acts to maintain a degree of isotropy by feeding the spanwise component, although the overall anisotropy is preserved due to the wall constraint. The other budget terms, such as dissipation $\varepsilon^+_{33}$, act as sinks and balance the budget.

\begin{figure}[ht!]
    \centering
    \includegraphics[width=0.95\linewidth]{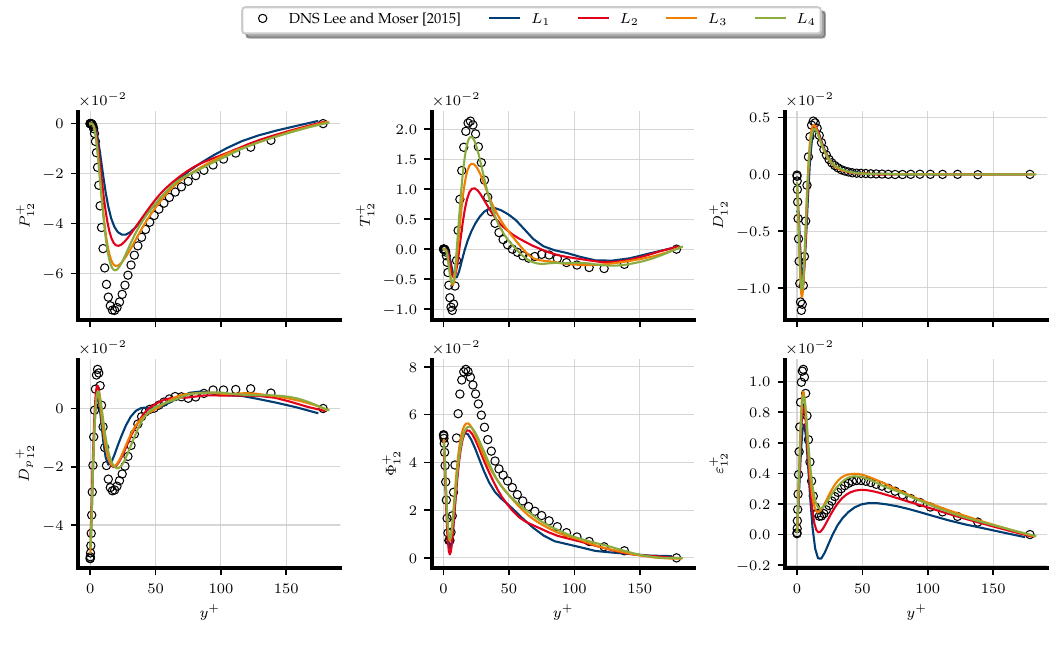}
    \caption{RSTE budget for shear stress component $\langle u^{\prime}_1 u^{\prime}_2 \rangle$.}
    \label{fig:chan_uv_RSTE}
\end{figure}

Figure \ref{fig:chan_uv_RSTE} shows the Reynolds shear stress budget for $R_{12}$, which is fundamental to understanding momentum transport. The production term $P^+_{12}$ arises from the interaction between the mean velocity gradient and the wall-normal Reynolds stress. This production is primarily balanced by the pressure-strain correlation $\Phi_{12}$, which acts to destroy the shear stress. The balance between these two terms is what eddy-viscosity models attempt to approximate with an algebraic relation. The finer meshes capture this balance more accurately, particularly in the near-wall region where viscous diffusion becomes important. As seen in previous results, the $12$ term of the RSTE burger follows similar trends where finer meshes are more accurate and simulation results predict DNS data with marginal accuracy.

\begin{figure}[ht!]
    \centering
    \includegraphics[width=0.95\linewidth]{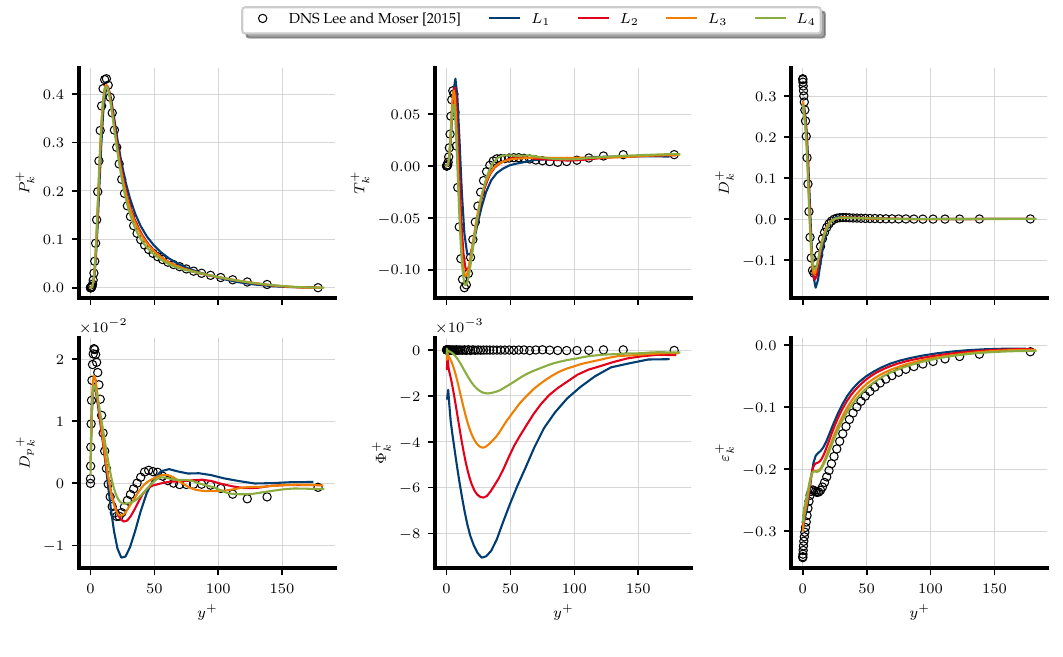}
    \caption{Comparison for the TKE budget.}
    \label{fig:chan_TKE_budget}
\end{figure}

The turbulent kinetic energy budget in Fig. \ref{fig:chan_TKE_budget} provides the overall energy balance for the turbulent fluctuations. The production term $P^+_k$ represents the energy extracted from the mean flow and is balanced by dissipation $\varepsilon^+$, which converts turbulent energy to heat. The remaining terms represent the spatial redistribution of the energy. In the log-law region, production and dissipation are in approximate equilibrium, a key assumption in many turbulence models. Near the wall, transport by viscous diffusion becomes a critical source of energy, while dissipation remains the primary sink. The finer meshes capture this balance more accurately, particularly in the near-wall region where the gradients are steepest.
Similarly, the pressure-strain terms are observed to converge towards zero as the mesh is refined.

\begin{figure}[ht!]
    \centering
    \includegraphics[width=0.95\linewidth]{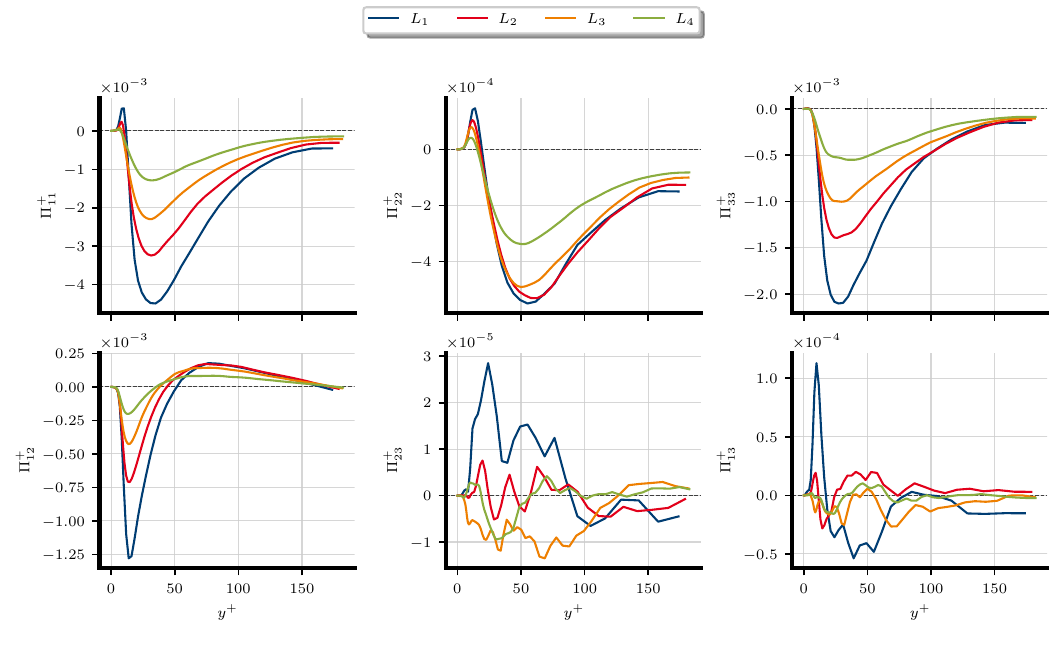}
    \caption{The SGS term $\Pi_{ij}$ from each of the RSTE budgets.}
    \label{fig:SGS_term}
\end{figure}

To end this section, the SGS contributions to each of the RSTE budgets are also considered.
These are shown in Fig.~\ref{fig:SGS_term}.
As expected, the SGS contributions converge towards zero as the mesh is refined, highlighting the diminishing contribution from unresolved fluctuations.
Additionally, the SGS contributions are seen to be small compared with the remaining terms in the RSTE budgets, which is especially evident on the finer meshes.

\subsection{Pipe Flow}
\label{subsec:pipe_flow}

To further illustrate the correct implementation of the RSTE budget, another canonical flow case is simulated in the same fashion as the previous channel flow results for different meshes.
The results are then compared against DNS reference data from \cite{el2013direct} of pipe flow at Re$_{\tau}=180$. Therefore, this section displays 1D-mean values of the first and second order statistics of the velocity---$\langle u \rangle$, $k$, $R_{ij}$---as well as the TKE budget, which is the trace of the RSTE budget.
As the trace is a scalar, it is an invariant and issues related to a change in coordinate-change are avoided. All values are shown in wall units.

\begin{figure}[ht!]
    \centering
    \includegraphics[width=0.95\textwidth]{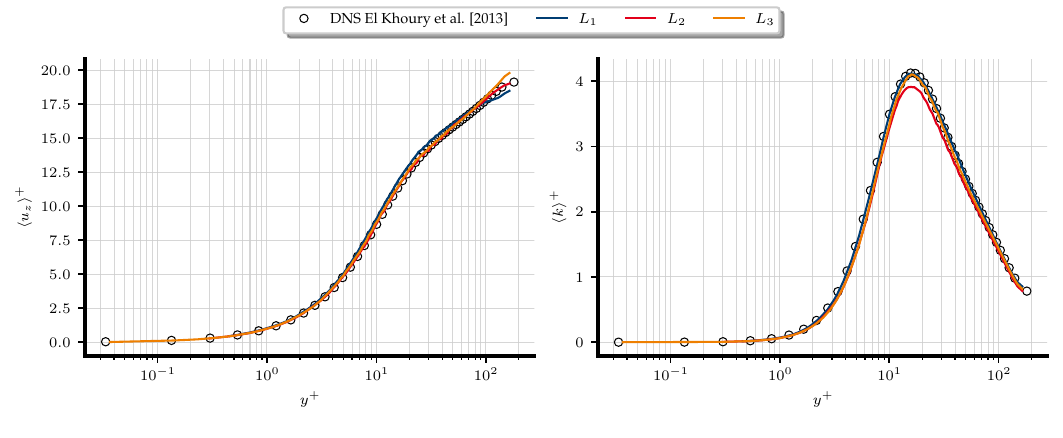}
    \caption{Streamwise velocity $\langle u_{z} \rangle$ and TKE $k$ profiles.}
    \label{fig:Fig1}
\end{figure}

Figure \ref{fig:Fig1} presents the velocity and TKE profiles for pipe flow, which can be directly compared with the channel flow results. The pipe flow exhibits similar characteristics to channel flow, with the mean velocity following the law of the wall in the logarithmic region. However, the pipe geometry introduces subtle differences mainly due to the curvature of the pipe walls, which is not present in the channel case.
The TKE profile shows a maximum in the buffer region, similar to channel flow, but this maximum location may differ slightly due to the curvature of the conduit. The finer meshes capture both the mean velocity and TKE profiles with improved accuracy, particularly in the near-wall region.

\begin{figure}[ht!]
    \centering
    \includegraphics[width=0.95\textwidth]{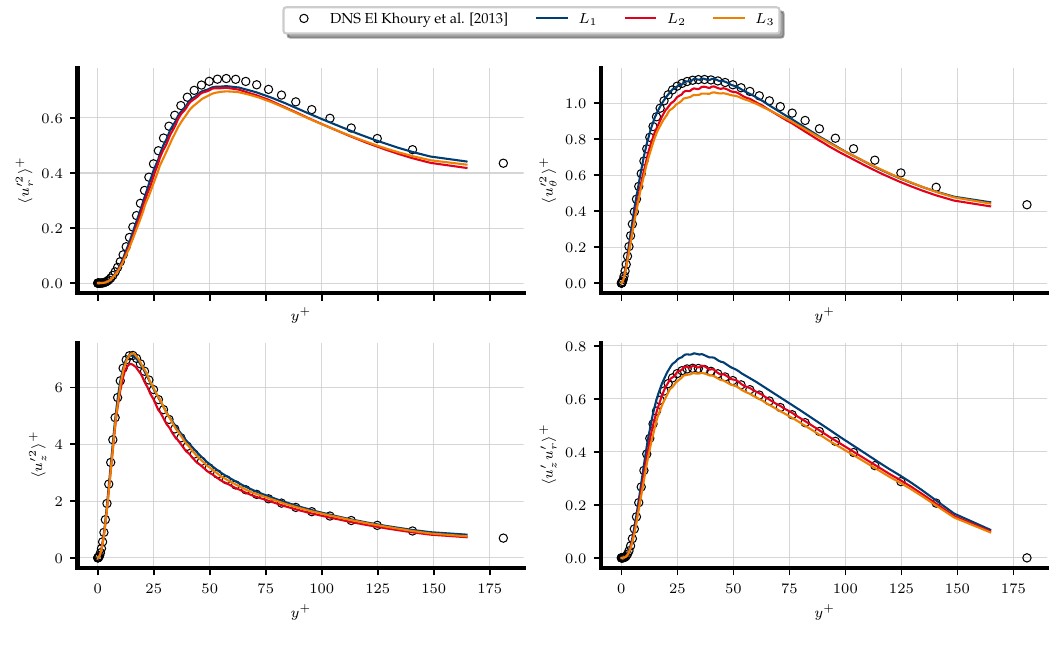}
    \caption{Profiles of $R_{ij}$, specifically components $zz$, $rr$, $\theta\theta$, and $zr$.}
    \label{fig:Fig2}
\end{figure}

The Reynolds stress components in pipe flow, shown in Fig. \ref{fig:Fig2}, demonstrate the same general trends as channel flow, with the streamwise component $R_{zz}$ being the largest and the circumferential component $R_{\theta\theta}$ being the smallest. The wall-normal radial component $R_{rr}$ and the radial shear stress component $R_{zr}$ show similar behaviour to channel flow, though the coordinate system differences (cylindrical vs. Cartesian) may introduce subtle variations in the interpretation of these components. The finer meshes provide improved prediction of the Reynolds stress profiles, particularly in the buffer region where the stresses peak and the gradients are significant.

\begin{figure}[ht!]
    \centering
    \includegraphics[width=0.95\textwidth]{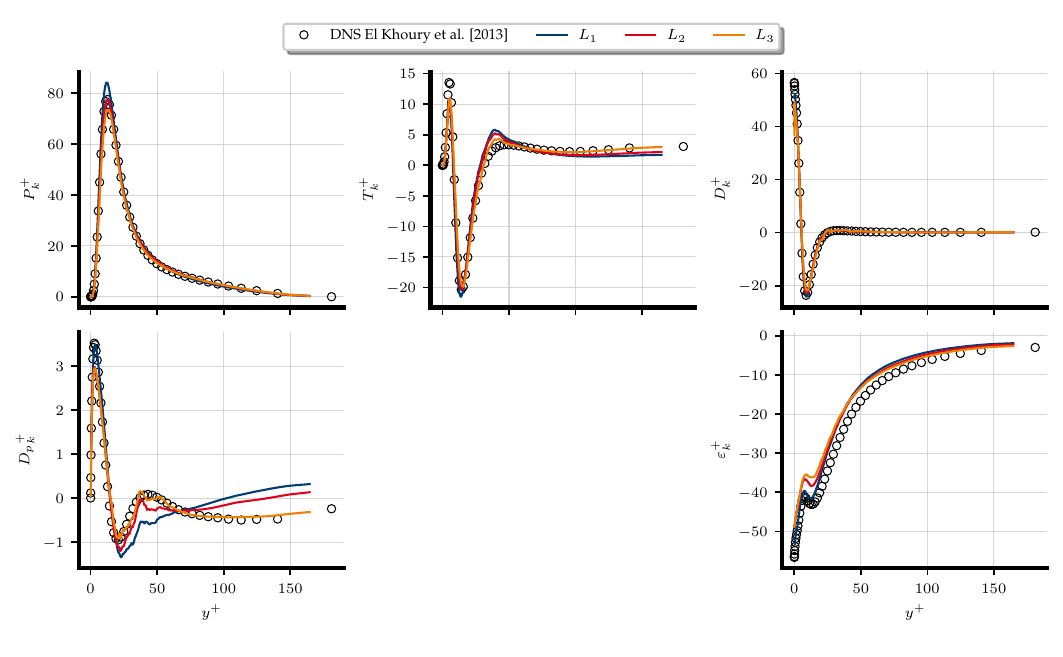}
    \caption{TKE budget of the pipe-flow simulation.}
    \label{fig:Fig3}
\end{figure}

Figure \ref{fig:Fig3} displays the trace of the RSTE budget components, which is equivalent to the TKE budget. As the trace of a tensor is an invariant, this quantity is independent of the coordinate system used. This makes it an excellent tool for validation, as it confirms that the fundamental energy balances of the implementation are correct, irrespective of the geometric complexities introduced by the pipe's cylindrical coordinate system. The TKE budget shows the overall energy balance for the turbulent fluctuations, with production balancing dissipation in the log-law region and viscous effects dominating near the wall.
The remaining terms act to redistribute the energy in space.
As seen in previous cases, the finer meshes capture this balance more accurately, demonstrating the correct implementation.

\section{Conclusion}
\label{sec:conclusions}

In this study, a comprehensive methodology for calculating the complete resolved RSTE budget from LES within the OpenFOAM framework has been successfully implemented and validated. The implementation relies on standard OpenFOAM utilities and post-processing functions to compute all terms of the resolved RSTE budget in Cartesian coordinates, including convection, production, turbulent transport, pressure-diffusion, pressure-strain, viscous diffusion, viscous dissipation, and, finally, the SGS contribution.

The fidelity of the implementation was rigorously assessed by comparing results against DNS data for two canonical wall-bounded turbulent flows: channel flow and pipe flow, both at a friction Reynolds number of Re$_{\tau}=180$. For the channel flow case, the predicted mean velocity, TKE, and Reynolds stress profiles demonstrated reasonable agreement with DNS data. A mesh refinement study confirmed that the LES results systematically converge towards the reference data. Furthermore, the individual budgets for the main components of the Reynolds stress tensor ($R_{11}$, $R_{22}$, $R_{33}$, and $R_{12}$) were also considered, and its was found that the implementation correctly reproduces the complex interplay of energy production, dissipation, and redistribution mechanisms that characterise near-wall turbulence. While minor discrepancies with DNS data were observed, these are consistent with the inherent limitations of the second-order-accurate finite volume numerics employed. The primary objective was the validation of the resolved RSTE budget library, not the replication of high-order spectral DNS results.
The pipe flow simulation further corroborates the correct implementation of the RSTE budget. By analysing the TKE budget---which is coordinate-invariant---it has been confirmed that the underlying energy balances are correctly computed, irrespective of the geometry. This result highlights the robustness of the developed RSTE budget tool.

In conclusion, this work provides the scientific and engineering community with a validated, open-source tool for conducting detailed RSTE budget analysis in OpenFOAM. This utility enables deeper insights into the physics of turbulence and serves as a valuable resource for the development and validation of advanced turbulence models. Future efforts could be directed towards optimising the memory footprint of the implementation and extending its capabilities to directly handle cylindrical and/or spherical coordinate systems, thereby broadening its applicability to more complex geometries.

\section*{Acknowledgements}

\noindent 
M.J.R and M.R. acknowledge financial support from Innovation Fund Denmark (IFD) under Grant No. 3130-00007B and Kamstrup A/S. 
C.H. and M.A. acknowledge funding from VILLUM FONDEN under Grant No. VIL57365. 
The authors would also like to acknowledge the EuroHPC Joint Undertaking for awarding this project access to the EuroHPC supercomputer LUMI, hosted by CSC (Finland) and the LUMI consortium through a EuroHPC Regular Access call. This work was also partially supported by the Danish e-Infrastructure Cooperation (DeiC) National HPC under grant number DeiC-AU-N5-2025130.


\authorcontributions{
Conceptualisation, M.J.R. and C.H.;
methodology, M.J.R. and C.H.;
software, M.J.R. and C.H.;
validation, M.J.R. and C.H.;
formal analysis, M.J.R., C.H., M.R. and M.A.;
investigation, M.J.R. and C.H.;
resources, M.R. and M.A.;
data curation, M.J.R. and C.H.;
writing---original draft preparation, M.J.R. and C.H.;
writing---review and editing, M.R. and M.A.;
visualisation, M.J.R. and C.H.;
supervision, M.R. and M.A.;
project administration, M.J.R., C.H., M.R. and M.A.;
funding acquisition, M.R. and M.A.
All authors have read and agreed to the published version of the manuscript.
}

\bibliographystyle{unsrt}  
\bibliography{references}  

\end{document}